\documentclass[journal,a4paper]{IEEEtran}

\usepackage{cite}
\usepackage[pdftex]{graphicx}
\usepackage{amsmath,amssymb}
\usepackage{url}

\usepackage{acro}
\usepackage{csquotes}
\usepackage{booktabs}
\usepackage{xspace}

\usepackage{hyperref}
\usepackage{cleveref}

\usepackage{subcaption}

\usepackage[rgb]{xcolor}
\usepackage{tikz}
\usepackage{pgfplots}
\usetikzlibrary{calc}
\pgfplotsset{compat=1.18}
\usetikzlibrary{patterns}

\pgfplotsset{
    tick label style = {font=\footnotesize},
    label style = {font=\footnotesize},
    legend style = {font=\footnotesize},
    title style = {font=\footnotesize}
}

\newcommand\sentinelii{Sentinel-2\xspace}

\DeclareAcronym{leo}{
	short={LEO},
	long={low Earth orbit}
}

\DeclareAcronym{meo}{
	short={MEO},
	long={medium Earth orbit}
}

\DeclareAcronym{geo}{
	short={GEO},
	long={geostationary}
}

\DeclareAcronym{eo}{
	short={EO},
	long={Earth observation}
}

\DeclareAcronym{cnn}{
	short={CNN},
	long={convolutional neural network}
}

\DeclareAcronym{iss}{
  short = ISS,
  long = International Space Station,
}

\DeclareAcronym{cots}{
  short = COTS,
  long = commercial off-the-shelf,
}

\DeclareAcronym{awgn}{
  short = AWGN,
  long = additive white Gaussian noise,
}

\DeclareAcronym{ldpc}{
  short = LDPC,
  long = low-density parity-check,
}

\DeclareAcronym{jscc}{
  short = JSCC,
  long = joint source-channel coding,
}

\DeclareAcronym{djscc}{
  short = {DJSCC},
  long = deep joint source-channel coding,
}

\DeclareAcronym{snr}{
  short = SNR,
  long = signal-to-noise ratio,
}

\DeclareAcronym{psnr}{
  short = PSNR,
  long = peak signal-to-noise ratio,
}

\DeclareAcronym{djscc-sat}{
  short = \textsc{DJSCC-Sat},
  long = {deep joint source-and-channel coding for small satellite applications}
}

\DeclareAcronym{adjscc-sat}{
  short = \textsc{$\mathcal{A}$DJSCC-Sat},
  long = {adaptable deep joint source-and-channel coding for small satellite applications}
}

\newcommand\MP{\ensuremath{\mathrm{MP}}\xspace}

\newcommand\jpegtwok{JPEG\,2000\xspace}

\DeclareAcronym{los}{
  short = {LOS},
  long = {line of sight}
}

\DeclareAcronym{relu}{
  short = ReLU,
  long = rectified linear unit,
}

\definecolor{plt1}  {Hsb}{131,.5,.80}
\definecolor{plt1-2}{Hsb}{141,.7,.70}
\definecolor{plt1-3}{Hsb}{151,.4,.60}
\definecolor{plt1-4}{Hsb}{161,.9,.50}
\definecolor{plt1-5}{Hsb}{171,.5,.40}

\definecolor{plt2}  {Hsb}{251,.5,.80}
\definecolor{plt2-2}{Hsb}{261,.7,.70}
\definecolor{plt2-3}{Hsb}{271,.4,.60}
\definecolor{plt2-4}{Hsb}{281,.9,.40}
\definecolor{plt2-5}{Hsb}{291,.5,.30}

\definecolor{plt3}  {Hsb}{311,.5,.80}
\definecolor{plt3-2}{Hsb}{321,.7,.70}
\definecolor{plt3-3}{Hsb}{331,.4,.60}
\definecolor{plt3-4}{Hsb}{341,.9,.50}
\definecolor{plt3-5}{Hsb}{351,.5,.40}

\definecolor{plt4}  {Hsb}{ 11,.5,.90}
\definecolor{plt4-2}{Hsb}{ 21,.7,.80}
\definecolor{plt4-3}{Hsb}{ 31,.3,.60}
\definecolor{plt4-4}{Hsb}{ 41,.9,.50}
\definecolor{plt4-5}{Hsb}{ 51,.5,.40}

\definecolor{plt5}  {Hsb}{ 71,.5,.80}
\definecolor{plt5-2}{Hsb}{ 81,.7,.70}
\definecolor{plt5-3}{Hsb}{ 91,.4,.60}
\definecolor{plt5-4}{Hsb}{101,.9,.50}
\definecolor{plt5-5}{Hsb}{111,.5,.40}

\definecolor{plt6}  {Hsb}{191,.5,.80}
\definecolor{plt6-2}{Hsb}{201,.7,.70}
\definecolor{plt6-3}{Hsb}{211,.4,.60}
\definecolor{plt6-4}{Hsb}{221,.9,.50}
\definecolor{plt6-5}{Hsb}{231,.5,.40}

\begin{document}

\title{Deep Joint Source-Channel Coding for Small Satellite Applications}

\author{Olga Kondrateva,
        Grace Li Zhang,
        Julian Zobel, 
        Björn Scheuermann, and
        Stefan Dietzel
\thanks{Olga Kondrateva, Grace Li Zhang, Julian Zobel, and Björn Scheuermann are with the Department of Electrical Engineering and Information Technology at the Technical University of Darmstadt, Germany.}%
\thanks{Stefan Dietzel is with Merantix Momentum GmbH, Berlin, Germany.}%
}

\maketitle

\begin{abstract}
Small satellites used for Earth observation generate vast amounts of high-dimensional data, but their operation in low Earth orbit creates a significant communication bottleneck due to limited contact times and harsh, varying channel conditions. While \ac{djscc} has emerged as a promising technique, its practical application to the complex satellite environment remains an open question. This paper presents a comprehensive \ac{djscc} framework tailored for satellite communications. We first establish a basic system, \acs{djscc-sat}, and integrate a realistic, multi-state statistical channel model to guide its training and evaluation. To overcome the impracticality of using separate models for every channel condition, we then introduce an adaptable architecture, \acs{adjscc-sat}, which leverages attention modules to allow a single neural network to adjust to a wide range of channel states with minimal overhead. Through extensive evaluation on Sentinel-2 multi-spectral data, we demonstrate that our adaptable approach achieves performance comparable to using multiple specialized networks while significantly reducing model storage requirements. Furthermore, the adaptable model shows enhanced robustness to channel estimation errors, outperforming the non-adaptable baseline. The proposed framework is a practical and efficient step toward deploying robust, adaptive \ac{djscc} systems for real-world satellite missions.
\end{abstract}

\begin{tikzpicture}[remember picture,overlay]
  \node[
    anchor=south west,
    text width=\textwidth,
    font=\sffamily\footnotesize
  ] at ($(current page.south west) + (0.5in,1)$) {%
  This work has been submitted to the IEEE for possible publication.\\Copyright may be transferred without notice, after which this version may no longer be accessible.
  };
\end{tikzpicture}

\acresetall

% ...

\IEEEpeerreviewmaketitle

\section{Introduction}

\IEEEPARstart{O}{ver} the last years, the number of small satellite missions has grown rapidly, with nanosatellites based on the CubeSat standard~\cite{cubesat2020} gaining particular popularity. Their key characteristics such as small size and the availability of \ac{cots} components have made access to space easier and more cost-effective~\cite{WOELLERT2011663}. This, in turn, has enabled a wide range of new applications across several important domains, including communication~\cite{saeed2019}, weather monitoring~\cite{7327032}, disaster management~\cite{barmpoutis2020}, and scientific research~\cite{BUDIANU201514, POGHOSYAN201759}.

Achieving state-of-the-art results in many of these domains requires high spatiotemporal or spectral resolution~\cite{9463743}. This need for high-quality data, coupled with ongoing advances in sensor technology, has created a severe communication bottleneck. The data volumes produced by modern sensors now significantly outpace the communication capabilities of small satellites. Operating in \ac{leo}, these satellites have only a few short contact windows with ground stations each day~\cite{vasisht2021}. In addition, unreliable and highly variable channel conditions due to atmospheric losses~\cite{nasa-report}, along with strict physical constraints on antenna size and power~\cite{saeed2019}, further intensify the problem. Consequently, the growth in data generation rates continues to outpace any increases in available download speeds~\cite{furano2020}.

To address this challenge, this work explores a joint approach that brings together networking and data science perspectives by developing an advanced \ac{djscc} framework. Traditionally, source and channel coding are treated separately. Although Shannon's separation theorem~\cite{cover1991elements} states this can be optimal, its assumption of an infinite code block length is not realistic. In real-world scenarios with finite lengths, joint optimization has the potential to achieve better communication performance~\cite{6408177}. Recently, learning-based joint source-channel coding schemes have been proposed~\cite{Bourtsoulatze2019}, but their suitability for the complex and rapidly varying satellite channel remains an open question.

Our work addresses this gap in two main stages. First, we introduce and evaluate a basic deep joint source and channel coding (DJSCC) system using neural networks specifically for the transmission of satellite images, which we term \ac{djscc-sat} \cite{lcn2023}. This initial step demonstrates the viability of applying end-to-end learning to this specific domain.

Second, to make the system practical for real-world deployment, we enhance this basic design to handle the complexities of the satellite channel. A major limitation of prior work has been its reliance on simplified channel models, like \ac{awgn}, which do not capture the dynamic nature of satellite links. To foster a more robust design, we integrate a realistic statistical channel model by Fontán et al.~\cite{966585}, which accounts for multiple shadowing states (e.g., \ac{los}, shadow, and deep shadow). However, this necessary step toward a more realistic system introduces a critical scalability bottleneck, as training a separate model for all possible channel conditions becomes computationally infeasible due to the associated overhead.

To overcome this challenge, we then introduce our adaptable \ac{djscc} architecture, \acs{adjscc-sat}, which enhances the baseline system with attention modules~\cite{arxiv2024}. This allows a single, compact neural network to be parameterized in real-time according to the current channel state. The result is a flexible and efficient solution capable of handling a wide range of channel conditions without the need for multiple separate models.

The present paper builds upon our preliminary work~\cite{lcn2023, arxiv2024, kondrateva-phd-2025}, cohesively integrating the basic~\cite{lcn2023} and the adaptable~\cite{arxiv2024, kondrateva-phd-2025} architectures into a unified framework and presenting the complete research arc in a single expanded methodology. 

The main contributions compared to our previously published peer-reviewed results are:
\begin{itemize}
    \item The design of the baseline \acs{djscc-sat} system for multi-spectral satellite imagery, now rigorously evaluated against a realistic statistical channel model that accounts for multiple shadowing states.
    \item The development of an adaptable architecture, \acs{adjscc-sat}, which integrates attention modules to allow a single neural network to adjust to a wide range of channel conditions with minimal overhead.
    \item A comprehensive performance evaluation on Sentinel-2 \cite{sentinel2} data that directly compares the adaptable and non-adaptable architectures and analyzes their robustness to channel estimation errors.
\end{itemize}


The remainder of this paper is organized as follows. \Cref{sec:related_work} reviews related work. \Cref{sec:system_model} details the system model and communication architectures. \Cref{sec:methodology} presents our methodology, from the basic to the adaptable \ac{djscc} system. \Cref{sec:evaluation} provides a comprehensive performance evaluation, and \Cref{sec:conclusion} concludes the paper.

\section{Related Work}
\label{sec:related_work}

This section reviews existing literature on \ac{jscc}, encompassing both traditional methods and those leveraging neural networks.

The field of \ac{jscc} has seen substantial research activity in recent years. A range of cross-layer optimization techniques were developed to jointly coordinate the selection of parameters across source coding, channel coding, and modulation. These methods were explored in diverse areas including terrestrial wireless networks~\cite{Wei2004, Appadwedula1998, Jianfei2000} and even deep-space links~\cite{Bursalioglu2011}. A significant barrier to their widespread adoption, however, was the prohibitive computational cost.

Deep \ac{jscc}, initially proposed by Bourtsoulatze et al.~\cite{Bourtsoulatze2019}, proved to be a promising solution, demonstrating performance comparable to state-of-the-art compression techniques while avoiding the complexity of earlier methods. Subsequent research has rapidly expanded its capabilities, enabling features like orthogonal frequency division multiplexing~\cite{yang2021}, progressive transmission~\cite{Kurka2021}, adaptive rate control~\cite{yang2021b}, use of non-differentiable channel models~\cite{Aoudia2019}, finite channel symbol alphabets (prior research used unrestricted complex values)~\cite{Tung2022},  and correlated data sources~\cite{Xuan2021}. However, two notable gaps remain: these approaches are typically evaluated on standard image datasets rather than specific satellite data, and they employ simplistic channel models ill-suited for the dynamic satellite environment.

Deep-learning-based methods for source compression alone have demonstrated superior performance over traditional techniques, particularly in preserving quality at high compression ratios~\cite{Hu2022}. This trend has extended to the satellite domain, with several notable contributions. Kong et al. proposed techniques for processing spectral and spatial image content in parallel~\cite{Kong2020} and for reducing redundancy using residual networks~\cite{Kong2020b}. For onboard applications, De Oliveira et al. introduced a lightweight variational auto-encoder~\cite{AlvesdeOliveira2021} and later combined it with onboard de-noising capabilities~\cite{9690871}. A key difference to our work, however, is that these approaches are purely focused on source compression, whereas our approach integrates this step with channel coding and modulation for end-to-end transmission.

Our focus on small satellite communication necessitates the use of appropriate channel models. While various technologies like visible light communication (VLC)~\cite{nakajima2012} and laser~\cite{welle2018} are being explored, radio-frequency (RF) links are still the most common. RF channel modeling has evolved from static models, which describe the link with a single distribution, to dynamic ones. Static approaches, such as those by Loo~\cite{1623307} and Corazza and Vatalaro~\cite{Corazza1994ASM} and their extensions~\cite{596315, 661055}, are unable to capture time-varying conditions.

Dynamic models overcome this by using state-based transitions, often leveraging Markov chains~\cite{saeed2019}, to represent changing propagation environments. Several such models exist, including those developed by Scalise et al.~\cite{4151152}, Nikolaidis et al.~\cite{7779114}, and Salamanca et al.~\cite{8693582}. For our approach, we selected the dynamic model by Fontan et al.~\cite{966585}. The explicit consideration of multiple shadowing states, combined with validation accross diverse frequencies, environments, and elevation angles, makes this model suitable for training and evaluating our proposed \ac{djscc} system. A comprehensive overview of other models can be found in Saeed et al.~\cite{saeed2019}.

\section{System Model and Communication Architecture}
\label{sec:system_model}

Our system model considers a single \ac{leo} satellite performing Earth observation, which must transmit its captured multispectral images to a ground station. The communication link is bandwidth-constrained, creating a bottleneck where the rate of data generation surpasses the available transmission rate, thereby requiring compression. For a practical application context, this work uses data from the Sentinel-2 mission, operated by the European Space Agency (ESA).

\subsection{Earth Observation Data and Onboard Processing}
\label{sub:eo_and_onboard}

\begin{figure*}
  \centering
  \begin{subfigure}{.3\linewidth}
    \includegraphics[width=\linewidth]{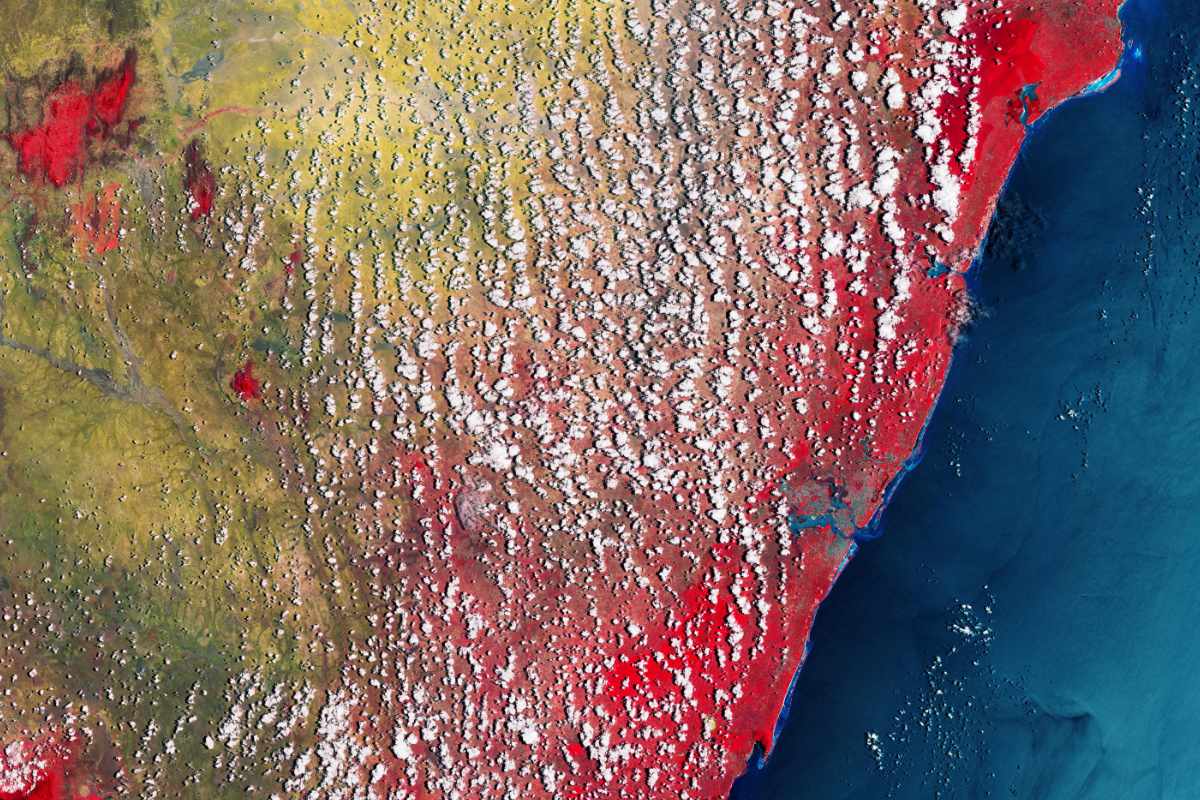}
    \caption{Southeast Kenya}
    \label{fig:sentinel_kenya}
  \end{subfigure}
  \hfill
  \begin{subfigure}{.3\linewidth}
    \includegraphics[width=\linewidth]{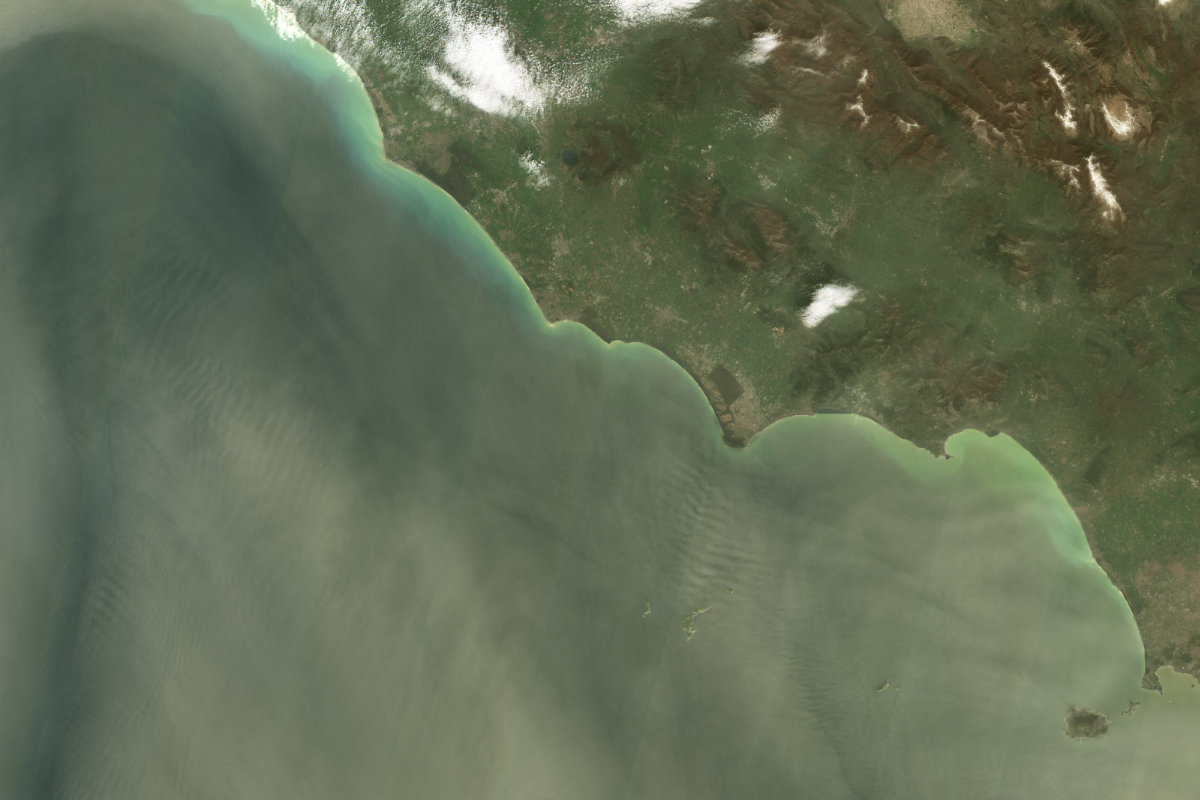}
    \caption{Sahara dust}
    \label{fig:sentinel_sahara}
  \end{subfigure}
  \hfill
  \begin{subfigure}{.3\linewidth}
    \includegraphics[width=\linewidth]{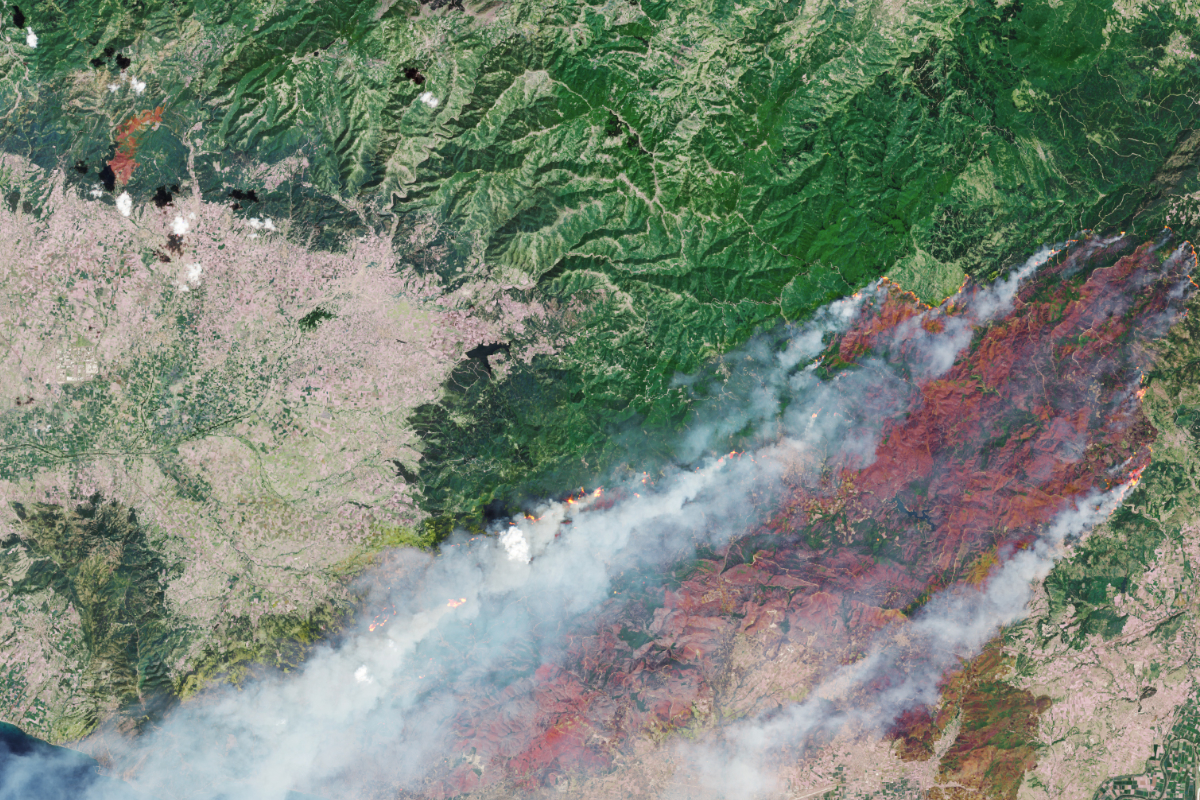}
    \caption{Wildfires in Greece}
    \label{fig:sentinel_greece}
  \end{subfigure}
  \caption{Example images from the \sentinelii mission demonstrating various Earth observation use cases. (Credit: processed by ESA, CC BY-SA 3.0 IGO)}
  \label{fig:sentinelii}
\end{figure*}

Sentinel-2 is an Earth observation mission from the Copernicus program,  which consists of a constellation of \ac{leo} satellites \cite{sentinel2}. Its satellites capture image data with a relatively high resolution of $10$\,m to $60$\,m, using a multi-spectral instrument to acquire images across 13 bands, including not only visible light but also near-infrared and short-wave infrared spectra. This results in high-dimensional image data that is valuable for a variety of use cases, from agricultural monitoring to disaster management. For instance, analyzing near-infrared channels allows for the assessment of vegetation health, while other bands can be used to monitor wildfires or dust storms, as shown in \Cref{fig:sentinelii}. By enabling more efficient and robust image transmission, the timeliness and utility of such data can be significantly improved.

The deployment of neural networks for such onboard processing tasks is increasingly feasible. The small size of satellites limits their energy budget, making low-power hardware a necessity. Recent studies and missions have successfully demonstrated the use of power-efficient processors for complex tasks onboard small satellites. For example, the Intel Movidius Myriad 2 and STM32 Microcontrollers have been used for star identification with a power envelope of approximately 1\,W \cite{s20216250}. Further, machine learning models like VGG19 and ResNet50 have been successfully evaluated on the \ac{iss} using Qualcomm and Intel Movidius processors \cite{9884906}, and the $\Phi$-Sat-1 mission has demonstrated onboard cloud detection, proving the feasibility of deploying AI at the edge in space \cite{9600851}.

\subsection{Separate vs. Joint Coding Architectures}
\label{sub:separate_vs_joint}

\begin{figure}
    \centering
    \includegraphics[width=.8\columnwidth]{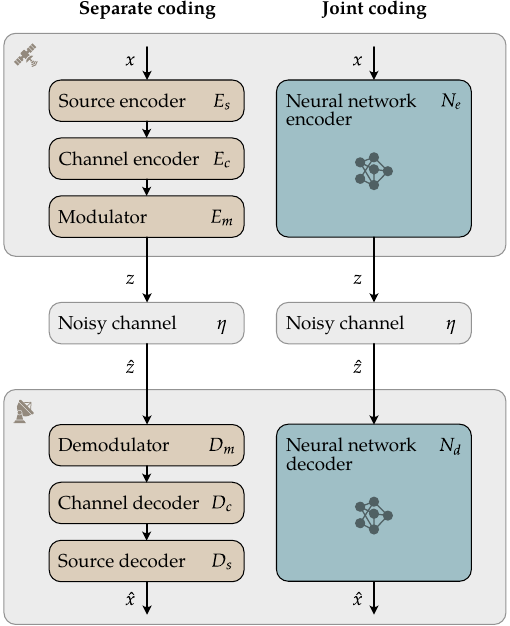}
    \caption{Comparison between the traditional communication model using separate coding and our joint coding approach.}
    \label{fig:jscc-overview}
\end{figure}

As illustrated in \Cref{fig:jscc-overview} (left), a traditional communication system treats source coding, channel coding, and modulation as separate components. For an input image $x \in \mathbb{R}^n$, the source encoder ($E_s$) first applies compression (e.g., \jpegtwok~\cite{sentinel-2-user-handbook}). Subsequently, the channel encoder ($E_c$) adds redundancy for error correction (e.g., using \ac{ldpc} codes). Finally, the modulator ($E_m$) converts the output into complex-valued samples $z \in \mathbb{C}^k$ suitable for transmission.

The ground station's receiver reverses this process, using a demodulator ($D_m$), channel decoder ($D_c$), and source decoder ($D_s$) to produce an estimate $\hat{x}$ of the original image from the received signal $\hat{z}$. Our work deviates from this modular paradigm.

As shown in \Cref{fig:jscc-overview} (right), we replace the separate components with a unified encoder ($N_e$) and decoder ($N_d$). These neural networks are trained end-to-end to learn an optimal mapping from the image $x$ to its channel representation $z$ and from the received signal $\hat{z}$ back to the reconstructed image $\hat{x}$.

The motivation for this joint design stems from the practical limitations of separate coding schemes. While Shannon's separation theorem proves optimality under ideal assumptions, these often do not hold in reality, leading to performance that can degrade significantly when there is a discrepancy between actual and expected channel conditions. This is particularly relevant for satellite links, which are subject to constant variation caused by movement, weather, and interference. In contrast, an end-to-end trained joint system can learn to be inherently robust to a wide range of channel conditions~\cite{Bourtsoulatze2019}.

\section{Joint Source-Channel Coding for Small Satellite Applications}
\label{sec:methodology}

\subsection{Basic \acs{jscc} Architecture (\acs{djscc-sat})}
\label{subsec:jscc_sat_arch}

Drawing on the considerations from \Cref{sub:separate_vs_joint}, we design our basic architecture, \acs{djscc-sat}, which is specifically tailored for the satellite communication domain. It utilizes an encoder-decoder structure where the encoder network jointly performs source coding, channel coding, and modulation. It directly maps an input image $x \in \mathbb{R}^n$ to a set of channel symbols $z \in \mathbb{C}^k$. The model's compression is controlled by the ratio $k/n$ (where $k < n$). Correspondingly, the decoder network is designed to reconstruct the image $\hat{x} \in \mathbb{R}^n$ from the received, and potentially noisy, symbol vector $\hat{z} \in \mathbb{C}^k$.

\begin{figure}
  \includegraphics[width=\linewidth]{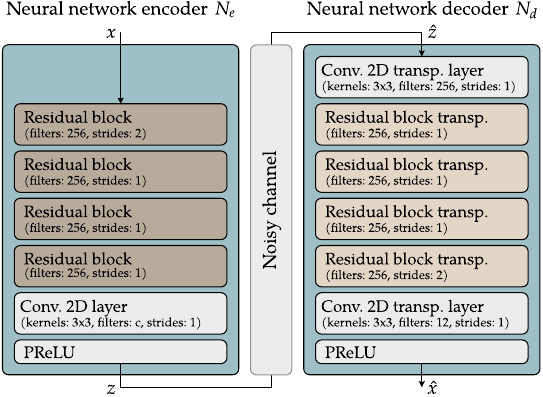}
  \caption{Encoder-decoder neural network architecture overview for the basic \ac{djscc-sat}.}
  \label{fig:model}
\end{figure}

\begin{figure}
  \centering

  \begin{subfigure}{.48\linewidth}
    \includegraphics[width=\linewidth]{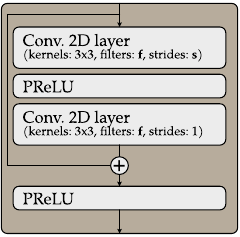}
    \caption{Residual block $(f, s)$}
  \end{subfigure}
  \hfill
  \begin{subfigure}{.48\linewidth}
    \includegraphics[width=\linewidth]{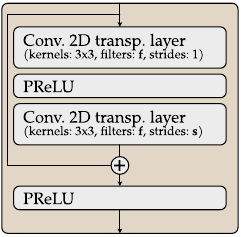}
    \caption{Residual block transp. $(f, s)$}
  \end{subfigure}

  \caption{Residual block architectures used in the \ac{djscc-sat} encoder and decoder.}
  \label{fig:residual}
\end{figure}

As shown in \Cref{fig:model,fig:residual}, our proposed architecture builds upon the ResNet model~\cite{resnet}, a common choice for image classification tasks. To produce channel-coded symbols, we modify the standard ResNet design by replacing its final dense layers -- typically used for classification -- with a new layer structure tailored for joint encoding. For the decoding process, a symmetrically reversed architecture is utilized.

The encoder is composed of four residual blocks, each containing 256 filters with a $3 \times 3$ kernel size. Its terminal layer is a convolutional layer with $c$ filters, where $c$ is a function of the desired compression ratio $k/n$. For the activation function, we utilize the parameterized rectified linear unit (PReLU), a generalization of ReLU that incorporates a learnable parameter to enhance predictive performance. The encoder's final output is a vector $\tilde{z}$ containing $k$ complex-valued channel symbols. To adhere to the average transmit power constraint $P$, the resulting vector $\tilde{z}$ is then normalized:
\begin{equation}
  z = \sqrt{kP}\frac{\tilde{z}}{\sqrt{\tilde{z}^{*} \tilde{z}}},
\end{equation}
where $\tilde{z}^{*}$ represents the conjugate transpose of $\tilde{z}$. This average transmit power constraint $P$ can be interpreted as the upper bound on the average power of the transmitted signal.

The decoder mirrors the encoder's structure. It is composed of a convolutional transpose layer, four residual transpose blocks, a second convolutional transpose layer, and a PReLU activation function. Its function is to process the received symbols $\hat{z}$ and produce a restored approximation of the original image, $\hat{x}$.

During training, we use the average mean squared error (MSE) as the loss function, which is defined as:
\begin{equation}
  \mathcal{L} = \mathrm{MSE} = \frac{1}{N} \sum_{i=1}^{N} d \bigl(x_i, \hat{x}_i\bigr),
\end{equation}
where $N$ is the number of samples and $d(x, \hat{x}) = ||x - \hat{x} ||^2$ denotes the MSE distortion. For evaluation, the \ac{psnr} is used as a metric to determine the quality of the reconstructed image:
\begin{equation}
  \mathrm{PSNR} = 10 \log_{10}\frac{\mathrm{MAX}^2}{\mathrm{MSE}},
\end{equation}
where $\mathrm{MAX}$ is the maximum possible pixel value. Intuitively, \ac{psnr} expresses the ratio between the maximum possible signal value and the distorting noise that reduces its quality.

\subsection{Satellite Channel Modeling}
\label{subsec:channel_model}

A critical component for training our system is a realistic simulation of the satellite-to-ground link. To achieve this, we insert a non-trainable layer between the encoder and decoder that models the channel. This layer simulates two primary characteristics: complex channel fading and additive noise, computing the received symbols $\hat{z}$ as follows:
\begin{equation}
  \hat{z} = zh + n,
\end{equation}
where $z$ are the transmitted symbols from the encoder, $h$ is the complex channel gain representing fading effects, and $n$ is the complex additive white Gaussian noise. In the following, we explain how $h$ and $n$ are calculated.

\subsubsection{Channel Gain}

To account for specific satellite channel conditions, we compute the channel gain $h$ using the statistical channel model by Fontan et al.~\cite{966585}. Since satellite channel characteristics heavily depend on shadowing conditions (e.g., due to buildings or foliage), they cannot be accurately modeled by a single distribution. Instead, the model introduces multiple states describing different degrees of shadowing, with transitions between them modeled by a three-state Markov chain. The states are:
\begin{itemize}
  \item \textbf{\acs{los} (line of sight)}: No shadowing.
  \item \textbf{Shadow}: Moderate shadowing conditions.
  \item \textbf{Deep shadow}: Heavy shadowing conditions.
\end{itemize}
Within each state, the channel is modeled using the Loo distribution~\cite{1623307}, which describes the received signal as a sum of a log-normally distributed direct component and a Rayleigh-distributed multipath component. The distribution is parameterized by the mean $\alpha$ and standard deviation $\psi$ of the direct signal, and the average power of the multipath component, \MP. The specific values for $\alpha$, $\psi$, and \MP are chosen depending on the environment (e.g., urban, suburban), the channel state, and the satellite's elevation angle, and can be obtained from statistical measurements~\cite{channel-params}. We simulate this distribution as described in~\cite{DBLP:journals/ijscn/Perez-FontanMMPMMR08} to generate the channel gain vector $h$.

\subsubsection{Additive Noise}
\label{subsubsec:additive_noise}

In order to compute the Gaussian noise vector $n$, we need to determine the \ac{snr} values that can be assumed.

The expected \ac{snr} can be computed as:
\begin{equation}
  \mathrm{SNR} = P_t + G_t + G_r - L - N,
\end{equation}
where $P_t$ is the transmitted power, $G_t$ and $G_r$ are the transmitter and receiver antenna gains, $L$ is the path loss, and $N$ is the thermal noise. All quantities are calculated in decibel. The $P_t$, $G_t$ and $G_r$ are input parameters depending on the particular equipment in use. $L$ and $N$ need to be estimated based on the assumed environment.

The path loss $L$ is determined from the Friis transmission formula:
\begin{equation}
  L = \frac{1}{G_tG_r}\Bigl(\frac{4\pi d f}{c}\Bigr)^2,
\end{equation}
where $d$ is the slant range, $f$ is the carrier frequency, and $c$ is the speed of light. The slant range $d$ is computed based on the satellite's orbit height $o$ and its elevation angle $\epsilon_{0}$. Assuming the satellite flying in an overhead trajectory, the slant range can then be computed using the following formula~\cite{7506756}:
\begin{equation}
  d = R_E\Biggl(\sqrt{\Bigl(\frac{o + R_E}{R_E}\Bigr)^2 - \cos^{2}\epsilon_{0}} - \sin\epsilon_{0}\Biggr),
\end{equation}
where $R_E$ is the radius of the Earth. 

The thermal noise $N$ is computed as $N=k \cdot T \cdot B$, where $k$ is Boltzmann's constant, $B$ is the channel bandwidth, and $T$ is the system noise temperature.

Finally, the computed \ac{snr} value is used to determine the noise power $\sigma^2$ for a given normalized signal power $P_{\mathrm{sig}}$:
\begin{equation}
  \sigma^2 = \frac{P_{\mathrm{sig}}}{2 \cdot 10^{\frac{\mathrm{SNR}}{10}}}.
\end{equation}
This allows us to generate the complex-valued noise vector $n$ using a normal distribution:
\begin{equation}
  n = \sigma \times \bigl[\mathcal{N}(0,1) + j \cdot \mathcal{N}(0,1) \bigr].
\end{equation}
This detailed channel model enables us to train and evaluate our system under a wide range of realistic satellite link conditions.

\subsection{Adaptable \acs{jscc} with Attention Modules (\ac{adjscc-sat})}
\label{subsec:adjscc_sat_arch}

Using the detailed channel model from \Cref{subsec:channel_model} results in a combinatorial increase in system complexity. A naive approach would require training and storing a separate network for each set of channel parameters ($\alpha$, $\psi$, \MP, and \ac{snr}), which is infeasible in practice. Our solution, which we call \acf{adjscc-sat}, is to augment the basic \ac{djscc-sat} architecture with attention modules. This enhancement provides a mechanism to parameterize a single, universal network based on the current channel state, thereby drastically reducing the system's complexity and overhead.

\begin{figure}
  \centering
  \includegraphics[width=\linewidth]{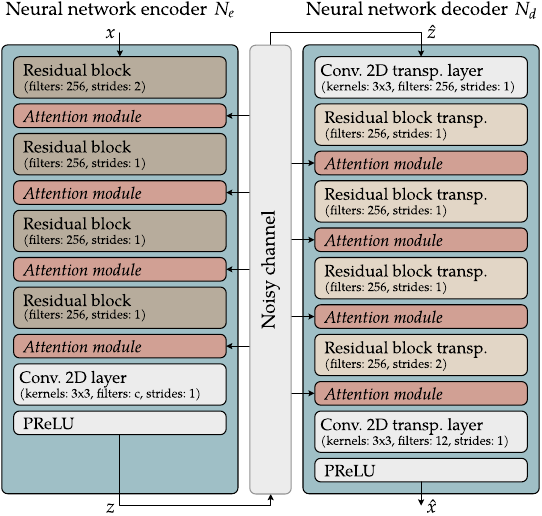}
  \caption{The \acs{adjscc-sat} architecture, enhancing the basic \ac{djscc-sat} network with attention modules after each residual and residual transpose block.}
  \label{fig:adjscc-arch}
\end{figure}

\begin{figure}
  \centering
  \includegraphics[width=.48\columnwidth]{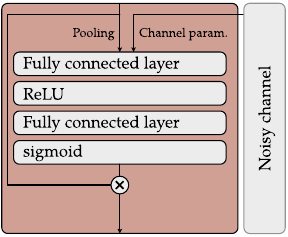}
  \caption{Structure of the attention module. Channel parameters are concatenated with pooled features to predict scaling factors, which are then applied to the original feature maps.}
  \label{fig:attention-module}
\end{figure}

Following the work of~\cite{wireless-attention-modules}, we insert an attention module after each of the four residual blocks in both the encoder and the decoder. The resulting architecture is shown in \Cref{fig:adjscc-arch}. The attention modules alter the feature weights of their preceding residual blocks to accommodate different channel conditions. 

The structure of an attention module is detailed in \Cref{fig:attention-module}. The process includes three key steps:
\begin{enumerate}
    \item \textbf{Context Extraction:} Global average pooling is applied to the output of the preceding residual block to extract global context information. This pooled output is then concatenated with a vector of variables representing the current channel parameters ($\alpha$, $\psi$, \MP, and \ac{snr} values).
    \item \textbf{Scaling Factor Prediction:} The concatenated vector is passed through a simple neural network consisting of two fully connected layers with \ac{relu} and sigmoid activation functions. This network predicts the appropriate scaling factors based on the channel condition parameters.
    \item \textbf{Attention Application:} Finally, the output of the preceding residual layer (before pooling) is multiplied element-wise with the predicted scaling factors, effectively re-weighting the feature maps to implement the attention-based scaling.
\end{enumerate}
This design allows a single, adaptable network to be parameterized for a wide range of channel conditions, providing a practical solution to the complexity introduced by realistic channel modeling. The trained \acs{adjscc-sat} network is then separated into its encoder component $N_e$ for use on the satellite and decoder component $N_d$ for use on the ground station.

\section{Performance Evaluation}
\label{sec:evaluation}

In this section, we evaluate our proposed methods. We first describe the experimental setup in \Cref{subsec:eval_setup}. We then analyze the performance of the basic deep joint source 
and channel coding (the \acs{djscc-sat} architecture), where a separate network is trained for each channel condition (\Cref{subsec:eval_jsscsat}). This serves as a baseline to demonstrate the general effectiveness of \ac{djscc} in this domain. Subsequently, in \Cref{subsec:eval_adjsccsat}, we evaluate our \acs{adjscc-sat} architecture, which uses attention modules, and compare its performance against the non-adaptable baseline. Finally, we assess the robustness of both approaches in case of channel estimation errors in \Cref{subsec:eval_mismatch}. For all scenarios, we compare the achieved image quality measured using \ac{psnr}, where higher values indicate better quality.

\subsection{Experimental Setup}
\label{subsec:eval_setup}

We evaluate two main \ac{djscc} approaches:
\begin{itemize}
    \item \textbf{\acs{djscc-sat}}: This approach uses the basic \ac{djscc} architecture described in \Cref{subsec:jscc_sat_arch}. A separate neural network is trained for each specific set of channel conditions (i.e., for each combination of environment, shadowing state, and elevation angle).
    \item \textbf{\acs{adjscc-sat}}: This is our proposed adaptable \ac{djscc} architecture described in \Cref{subsec:adjscc_sat_arch}, which uses attention modules, enabling a single neural network to adjust to varying channel conditions.
\end{itemize}

For training and evaluation, we utilized a dataset of \sentinelii multi-spectral images from the BigEarth archive~\cite{sumbul2019bigearthnet,Sumbul2021}, focusing on the region of Serbia during the summer. We first refined the dataset by removing images with heavy cloud cover and excluding the cloud-focused band 10, resulting in a set of 14,439 images. A series of preprocessing steps were then applied: all spectral bands were resized to a uniform resolution using cubic interpolation, and pixel values were normalized to a range between 0 and 1. Finally, the dataset was partitioned into training, validation, and test sets.

All models were implemented using Keras~\cite{keras} and Tensorflow~\cite{tensorflow} and trained using the Adam optimizer. We employed a batch size of 32 and an initial learning rate of $10^{-3}$, which was reduced to $10^{-4}$ after 500 epochs to facilitate fine-tuning.

To simulate the channel conditions detailed in \Cref{subsec:channel_model}, we utilized statistical measurements for $\alpha$, $\psi$, and \MP from~\cite{channel-params}. These parameters cover five distinct environments (\emph{open, suburban, intermediate tree shadow, heavy tree shadow,} and \emph{urban}) and three shadowing states (\emph{\acl{los}, shadow,} and \emph{deep shadow}), for satellite elevation angles between 40\textdegree{} and 80\textdegree{}. The corresponding \ac{snr} for each scenario was calculated as described in \Cref{subsubsec:additive_noise}, using the channel parameters listed in \Cref{tab:channel_parameters}.

\begin{table}
  \caption{Channel Parameters.}
  \label{tab:channel_parameters}
  \centering
	\begin{tabular}{ll}
		\toprule
    Parameter & Value \\
    \midrule
		Orbit height & 750\,km \\
		Carrier frequency & 2150\,MHz \\
    Transmitted power & 1\,W \\
    Satellite antenna gain & 6\,dBi \\
    Ground station antenna gain & 35\,dBi \\
		Receive channel bandwidth & 750\,kHz \\
		Noise figure & 2\,dB \\
		\bottomrule
	\end{tabular}
\end{table}

\subsection{Performance of Basic \acs{djscc-sat}}
\label{subsec:eval_jsscsat}

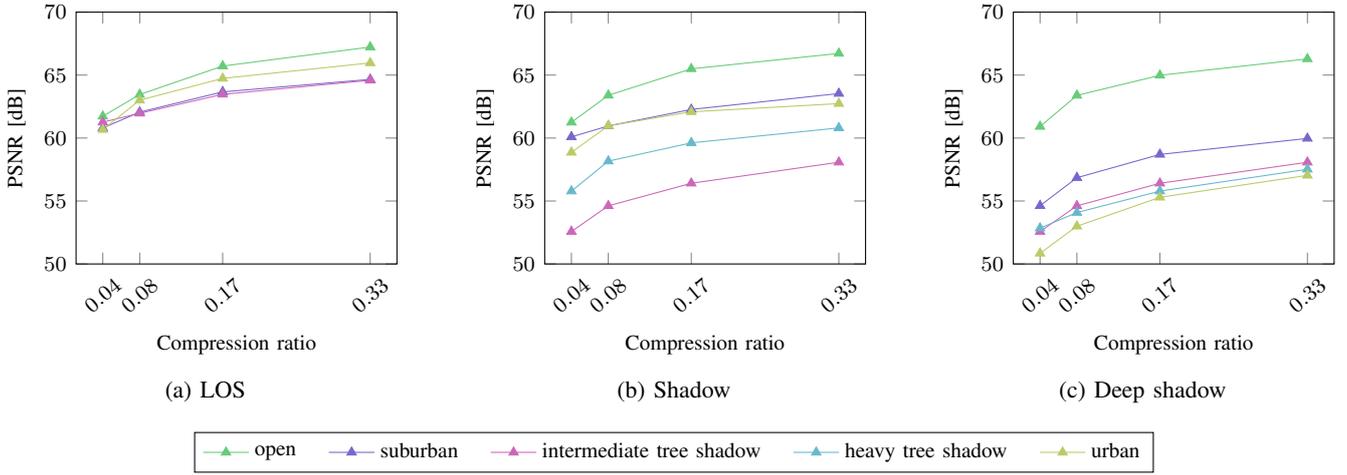
\begin{figure*}
  \begin{subfigure}{.32\linewidth}
    \centering
    \begin{tikzpicture}
    \begin{axis}[
        width=\linewidth,
        height=14em,
        xlabel = {Compression ratio},
        ylabel = {PSNR [dB]},
        ymin = 50,
        ymax = 70,
        xtick = data,
        x tick label style={
            rotate=40,
            /pgf/number format/fixed,
            /pgf/number format/precision=2,
        },
        ]

        \addplot[mark=triangle*,plt1] 
        plot 
        table[x expr=\thisrowno{0}, y expr=\thisrowno{1}, col sep=space] {plots/no_AF_40_open_los_los};

        \addplot[mark=triangle*,plt2] 
        plot 
        table[x expr=\thisrowno{0}, y expr=\thisrowno{1}, col sep=space] {plots/no_AF_40_suburban_los_los};

        \addplot[mark=triangle*,plt3] 
        plot 
        table[x expr=\thisrowno{0}, y expr=\thisrowno{1}, col sep=space] {plots/no_AF_40_intermediate_tree_shadow_los_los};
        
        \addplot[mark=triangle*,plt5] 
        plot 
        table[x expr=\thisrowno{0}, y expr=\thisrowno{1}, col sep=space] {plots/no_AF_40_urban_los_los};

    \end{axis}
\end{tikzpicture}
    \caption{LOS}
    \label{subfig:baseline_los}
  \end{subfigure}
  \hfill
  \begin{subfigure}{.32\linewidth}
    \centering
    \begin{tikzpicture}
    \begin{axis}[
        width=\linewidth,
        height=14em,
        xlabel = {Compression ratio},
        ylabel = {PSNR [dB]},
        ymin = 50,
        ymax = 70,
        xtick = data,
        x tick label style={
            rotate=40,
            /pgf/number format/fixed,
            /pgf/number format/precision=2,
        },
        legend cell align={left},
        legend columns=5,
		legend style={/tikz/every even column/.append style={column sep=1em}},
        legend to name={jscc-legendenv},
        ]

        \addplot[mark=triangle*,plt1] 
        plot 
        table[x expr=\thisrowno{0}, y expr=\thisrowno{1}, col sep=space] {plots/no_AF_40_open_shadow_shadow};

        \addplot[mark=triangle*,plt2] 
        plot 
        table[x expr=\thisrowno{0}, y expr=\thisrowno{1}, col sep=space] {plots/no_AF_40_suburban_shadow_shadow};

        \addplot[mark=triangle*,plt3] 
        plot 
        table[x expr=\thisrowno{0}, y expr=\thisrowno{1}, col sep=space] {plots/no_AF_40_intermediate_tree_shadow_deep_shadow_deep_shadow};
        
        \addplot[mark=triangle*,plt6] 
        plot 
        table[x expr=\thisrowno{0}, y expr=\thisrowno{1}, col sep=space] {plots/no_AF_40_heavy_tree_shadow_shadow_shadow};
        
        \addplot[mark=triangle*,plt5] 
        plot 
        table[x expr=\thisrowno{0}, y expr=\thisrowno{1}, col sep=space] {plots/no_AF_40_urban_shadow_shadow};

        \legend{
            open,
            suburban, 
            intermediate tree shadow,
            heavy tree shadow,
            urban
            }

    \end{axis}
\end{tikzpicture}
    \caption{Shadow}
    \label{subfig:baseline_shadow}
  \end{subfigure}
  \hfill
  \begin{subfigure}{.32\linewidth}
    \centering
    \begin{tikzpicture}
    \begin{axis}[
        width=\linewidth,
        height=14em,
        xlabel = {Compression ratio},
        ylabel = {PSNR [dB]},
        ymin = 50,
        ymax = 70,
        xtick = data,
        x tick label style={
            rotate=40,
            /pgf/number format/fixed,
            /pgf/number format/precision=2,
        },
        ]

        \addplot[mark=triangle*,plt1] 
        plot 
        table[x expr=\thisrowno{0}, y expr=\thisrowno{1}, col sep=space] {plots/no_AF_40_open_deep_shadow_deep_shadow};

        \addplot[mark=triangle*,plt2] 
        plot 
        table[x expr=\thisrowno{0}, y expr=\thisrowno{1}, col sep=space] {plots/no_AF_40_suburban_deep_shadow_deep_shadow};

        \addplot[mark=triangle*,plt3] 
        plot 
        table[x expr=\thisrowno{0}, y expr=\thisrowno{1}, col sep=space] {plots/no_AF_40_intermediate_tree_shadow_deep_shadow_deep_shadow};
        
        \addplot[mark=triangle*,plt6] 
        plot 
        table[x expr=\thisrowno{0}, y expr=\thisrowno{1}, col sep=space] {plots/no_AF_40_heavy_tree_shadow_deep_shadow_deep_shadow};
        
        \addplot[mark=triangle*,plt5] 
        plot 
        table[x expr=\thisrowno{0}, y expr=\thisrowno{1}, col sep=space] {plots/no_AF_40_urban_deep_shadow_deep_shadow};

    \end{axis}
\end{tikzpicture}
    \caption{Deep shadow}
    \label{subfig:baseline_deep_shadow}
  \end{subfigure}
  
  \vspace{1em}
  
  \centering
  {\hypersetup{hidelinks}\ref{jscc-legendenv}} 
  \caption{\ac{psnr} achieved by \acs{djscc-sat} for different environments, states, and compression ratios with 40\textdegree{} elevation angle.}
  \label{fig:different_scenes_40}
\end{figure*}

Next, we evaluate the performance of the basic \acs{djscc-sat} system, where a separate network is trained for each condition. \Cref{fig:different_scenes_40} shows the results for a 40\textdegree{} elevation angle across different environments and shadowing states. As expected, the best \ac{psnr} values are achieved in the open environment and the worst in more challenging environments, such as urban or intermediate tree shadow. The results also depend significantly on the shadowing conditions; for deep shadow (\Cref{subfig:baseline_deep_shadow}), the performance varies considerably between different environments, while the differences become less pronounced as shadowing conditions improve (\Cref{subfig:baseline_shadow,subfig:baseline_los}).

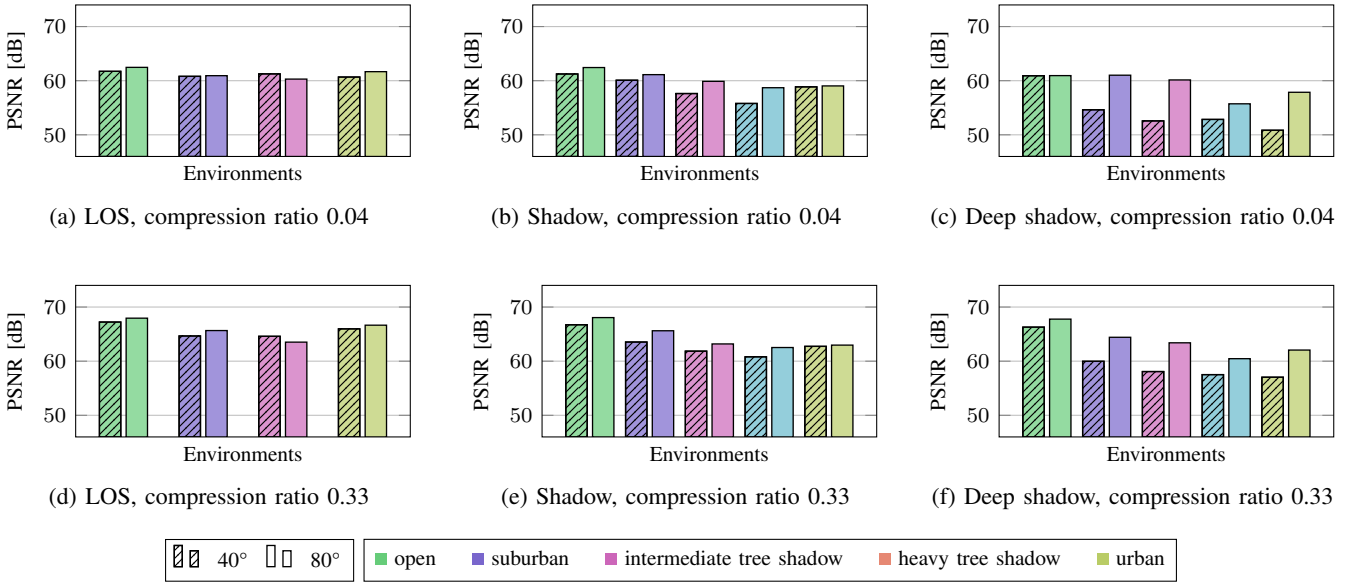
\begin{figure*}
  \begin{subfigure}{.33\linewidth}
    \centering
    \begin{tikzpicture}
    \begin{axis}[
      ybar,
      width=\linewidth,
      height=.6\linewidth,
      xmajorticks=false,
      every axis plot/.append style={
                bar width=8pt,
                bar shift=0pt,
                fill,
                line width=.5pt
              },
      ymajorgrids=true,
      ylabel={PSNR [dB]},
      ymin = 50,
        ymax = 70,
      enlargelimits=.2,
      xlabel={Environments},
      symbolic x coords ={%
          open,
          suburban,
          intermediate tree-shadow,
          urban,
      },
      legend cell align={left},
      legend style={column sep=5pt},
      legend to name={jscc-legendangles},
      legend columns=2,
      xticklabel=\empty,
      ]
      
      \legend{
        40\textdegree{},
        80\textdegree{}
      }

      \addplot[
        bar shift=-5pt,fill=white,postaction={
        pattern=north east lines, pattern color=black
    }] coordinates { (open, 0) };
      \addplot[bar shift=5pt,fill=white] coordinates { (open, 0) };

      \addplot[
        bar shift=-5pt,fill=plt1!70,postaction={
        pattern=north east lines, pattern color=black
    }] coordinates { (open, 61.73067188214748) };
      \addplot[bar shift=5pt,fill=plt1!70] coordinates { (open, 62.46548778477633) };

      \addplot[
      bar shift=-5pt,fill=plt2!70,postaction={
        pattern=north east lines, pattern color=black
    }] coordinates { (suburban, 60.805258435946925) };
      \addplot[bar shift=5pt,fill=plt2!70] coordinates { (suburban, 60.92828818309016) };
      
      \addplot[
        bar shift=-5pt,fill=plt3!70,postaction={
        pattern=north east lines, pattern color=black
    }] coordinates { (intermediate tree-shadow, 61.2610153054711) };
      \addplot[bar shift=5pt,fill=plt3!70] coordinates { (intermediate tree-shadow, 60.29798046606024) };
      
      \addplot[
        bar shift=-5pt,fill=plt5!70,postaction={
        pattern=north east lines, pattern color=black
    }] coordinates { (urban, 60.67812370860495) };
      \addplot[bar shift=5pt,fill=plt5!70] coordinates { (urban, 61.670784487418885) };
      
    \end{axis} 
  \end{tikzpicture}
    \caption{LOS, compression ratio 0.04}
    \label{subfig:base4080_los_04}
  \end{subfigure}%
  \hfill%
  \begin{subfigure}{.33\linewidth}
    \centering
    \begin{tikzpicture}
    \begin{axis}[
      ybar,
      width=\linewidth,
      height=.6\linewidth,
      xmajorticks=false,
      every axis plot/.append style={
                bar width=8pt,
                bar shift=0pt,
                fill,
                line width=.5pt
              },
      ymajorgrids=true,
      ylabel={PSNR [dB]},
      ymin = 50,
        ymax = 70,
      enlargelimits=.2,
      xlabel={Environments},
      symbolic x coords ={%
          open,
          suburban,
          intermediate tree-shadow,
          heavy tree-shadow,
          urban,
      },
      xticklabel=\empty,
      ]
  
      \addplot[
        bar shift=-5pt,fill=white,postaction={
        pattern=north east lines, pattern color=black
    }] coordinates { (open, 0) };
      \addplot[bar shift=5pt,fill=white] coordinates { (open, 0) };

      \addplot[
        bar shift=-5pt,fill=plt1!70,postaction={
        pattern=north east lines, pattern color=black
    }] coordinates { (open, 61.242307462063216) };
      \addplot[bar shift=5pt,fill=plt1!70] coordinates { (open, 62.43216132742652) };

      \addplot[
      bar shift=-5pt,fill=plt2!70,postaction={
        pattern=north east lines, pattern color=black
    }] coordinates { (suburban, 60.0867862998761) };
      \addplot[bar shift=5pt,fill=plt2!70] coordinates { (suburban, 61.12110295624692) };
      
      \addplot[
        bar shift=-5pt,fill=plt3!70,postaction={
        pattern=north east lines, pattern color=black
    }] coordinates { (intermediate tree-shadow, 57.62850182792831) };
      \addplot[bar shift=5pt,fill=plt3!70] coordinates { (intermediate tree-shadow, 59.88167207543309) };
      
      \addplot[
        bar shift=-5pt,fill=plt6!70,postaction={
        pattern=north east lines, pattern color=black
    }] coordinates { (heavy tree-shadow, 55.79653619027105) };
      \addplot[bar shift=5pt,fill=plt6!70] coordinates { (heavy tree-shadow, 58.7045005554619) };
      
      \addplot[
        bar shift=-5pt,fill=plt5!70,postaction={
        pattern=north east lines, pattern color=black
    }] coordinates { (urban, 58.862160183144205) };
      \addplot[bar shift=5pt,fill=plt5!70] coordinates { (urban, 59.03699565522086) };
      
    \end{axis} 
  \end{tikzpicture}
    \caption{Shadow, compression ratio 0.04}
    \label{subfig:base4080_shadow_04}
  \end{subfigure}
  \hfill%
  \begin{subfigure}{.33\linewidth}
    \centering
    \begin{tikzpicture}
    \begin{axis}[
      ybar,
      width=\linewidth,
      height=.6\linewidth,
      xmajorticks=false,
      every axis plot/.append style={
                bar width=8pt,
                bar shift=0pt,
                fill,
                line width=.5pt
              },
      ymajorgrids=true,
      ylabel={PSNR [dB]},
      ymin = 50,
        ymax = 70,
      enlargelimits=.2,
      xlabel={Environments},
      symbolic x coords ={%
          open,
          suburban,
          intermediate tree-shadow,
          heavy tree-shadow,
          urban,
      },
      xticklabel=\empty,
      ]
  
      \addplot[
        bar shift=-5pt,fill=white,postaction={
        pattern=north east lines, pattern color=black
    }] coordinates { (open, 0) };
      \addplot[bar shift=5pt,fill=white] coordinates { (open, 0) };

      \addplot[
        bar shift=-5pt,fill=plt1!70,postaction={
        pattern=north east lines, pattern color=black
    }] coordinates { (open, 60.915204772562106) };
      \addplot[bar shift=5pt,fill=plt1!70] coordinates { (open, 60.941951355437276) };

      \addplot[
      bar shift=-5pt,fill=plt2!70,postaction={
        pattern=north east lines, pattern color=black
    }] coordinates { (suburban, 54.626502435886835) };
      \addplot[bar shift=5pt,fill=plt2!70] coordinates { (suburban, 61.017142451016966) };
      
      \addplot[
        bar shift=-5pt,fill=plt3!70,postaction={
        pattern=north east lines, pattern color=black
    }] coordinates { (intermediate tree-shadow, 52.57299641191118) };
      \addplot[bar shift=5pt,fill=plt3!70] coordinates { (intermediate tree-shadow, 60.14165345147827) };
      
      \addplot[
        bar shift=-5pt,fill=plt6!70,postaction={
        pattern=north east lines, pattern color=black
    }] coordinates { (heavy tree-shadow, 52.84660367941581) };
      \addplot[bar shift=5pt,fill=plt6!70] coordinates { (heavy tree-shadow, 55.73117874454389) };
      
      \addplot[
        bar shift=-5pt,fill=plt5!70,postaction={
        pattern=north east lines, pattern color=black
    }] coordinates { (urban, 50.84575762551614) };
      \addplot[bar shift=5pt,fill=plt5!70] coordinates { (urban, 57.86157808541486) };
      
    \end{axis} 
  \end{tikzpicture}
    \caption{Deep shadow, compression ratio 0.04}
    \label{subfig:base4080_deep_shadow_04}
  \end{subfigure}%
  
  \vspace{2em}
  
  \begin{subfigure}{.33\linewidth}
    \centering
    \begin{tikzpicture}
    \begin{axis}[
      ybar,
      width=\linewidth,
      height=.6\linewidth,
      xmajorticks=false,
      every axis plot/.append style={
                bar width=8pt,
                bar shift=0pt,
                fill,
                line width=.5pt
              },
      ymajorgrids=true,
      ylabel={PSNR [dB]},
      ymin = 50,
        ymax = 70,
      enlargelimits=.2,
      xlabel={Environments},
      symbolic x coords ={%
          open,
          suburban,
          intermediate tree-shadow,
          urban,
      },
      xticklabel=\empty,
      ]

      \addplot[
        bar shift=-5pt,fill=white,postaction={
        pattern=north east lines, pattern color=black
    }] coordinates { (open, 0) };
      \addplot[bar shift=5pt,fill=white] coordinates { (open, 0) };

      \addplot[
        bar shift=-5pt,fill=plt1!70,postaction={
        pattern=north east lines, pattern color=black
    }] coordinates { (open, 67.22762412790338) };
      \addplot[bar shift=5pt,fill=plt1!70] coordinates { (open, 67.94286665070874) };

      \addplot[
      bar shift=-5pt,fill=plt2!70,postaction={
        pattern=north east lines, pattern color=black
    }] coordinates { (suburban, 64.65365825355423) };
      \addplot[bar shift=5pt,fill=plt2!70] coordinates { (suburban, 65.65526368893802) };
      
      \addplot[
        bar shift=-5pt,fill=plt3!70,postaction={
        pattern=north east lines, pattern color=black
    }] coordinates { (intermediate tree-shadow, 64.59177795781159) };
      \addplot[bar shift=5pt,fill=plt3!70] coordinates { (intermediate tree-shadow, 63.51533179513708) };
      
      \addplot[
        bar shift=-5pt,fill=plt5!70,postaction={
        pattern=north east lines, pattern color=black
    }] coordinates { (urban, 65.96625190247083) };
      \addplot[bar shift=5pt,fill=plt5!70] coordinates { (urban, 66.64719138769036) };
      
    \end{axis} 
  \end{tikzpicture}
    \caption{LOS, compression ratio 0.33}
    \label{subfig:base4080_los_33}
  \end{subfigure}
  \hfill%
  \begin{subfigure}{.33\linewidth}
    \centering
    \begin{tikzpicture}
    \begin{axis}[
      ybar,
      width=\linewidth,
      height=.6\linewidth,
      xmajorticks=false,
      every axis plot/.append style={
                bar width=8pt,
                bar shift=0pt,
                fill,
                line width=.5pt
              },
      ymajorgrids=true,
      ylabel={PSNR [dB]},
      ymin = 50,
        ymax = 70,
      enlargelimits=.2,
      xlabel={Environments},
      symbolic x coords ={%
          open,
          suburban,
          intermediate tree-shadow,
          heavy tree-shadow,
          urban,
      },
      xticklabel=\empty,
      ]

      \addplot[
        bar shift=-5pt,fill=white,postaction={
        pattern=north east lines, pattern color=black
    }] coordinates { (open, 0) };
      \addplot[bar shift=5pt,fill=white] coordinates { (open, 0) };

      \addplot[
        bar shift=-5pt,fill=plt1!70,postaction={
        pattern=north east lines, pattern color=black
    }] coordinates { (open, 66.72378601897488) };
      \addplot[bar shift=5pt,fill=plt1!70] coordinates { (open, 68.05071847613172) };

      \addplot[
      bar shift=-5pt,fill=plt2!70,postaction={
        pattern=north east lines, pattern color=black
    }] coordinates { (suburban, 63.5362946198049) };
      \addplot[bar shift=5pt,fill=plt2!70] coordinates { (suburban, 65.61713528754578) };
      
      \addplot[
        bar shift=-5pt,fill=plt3!70,postaction={
        pattern=north east lines, pattern color=black
    }] coordinates { (intermediate tree-shadow, 61.85098102603866) };
      \addplot[bar shift=5pt,fill=plt3!70] coordinates { (intermediate tree-shadow, 63.182928753310684) };
      
      \addplot[
        bar shift=-5pt,fill=plt6!70,postaction={
        pattern=north east lines, pattern color=black
    }] coordinates { (heavy tree-shadow, 60.80235241808859) };
      \addplot[bar shift=5pt,fill=plt6!70] coordinates { (heavy tree-shadow, 62.51972868765633) };
      
      \addplot[
        bar shift=-5pt,fill=plt5!70,postaction={
        pattern=north east lines, pattern color=black
    }] coordinates { (urban, 62.73554608081638) };
      \addplot[bar shift=5pt,fill=plt5!70] coordinates { (urban, 62.9654737199328) };
      
    \end{axis} 
  \end{tikzpicture}
    \caption{Shadow, compression ratio 0.33}
    \label{subfig:base4080_shadow_33}
  \end{subfigure}%
  \hfill%
  \begin{subfigure}{.33\linewidth}
    \centering
    \begin{tikzpicture}
    \begin{axis}[
      ybar,
      width=\linewidth,
      height=.6\linewidth,
      xmajorticks=false,
      every axis plot/.append style={
                bar width=8pt,
                bar shift=0pt,
                fill,
                line width=.5pt
              },
      ymajorgrids=true,
      ylabel={PSNR [dB]},
      ymin = 50,
        ymax = 70,
      enlargelimits=.2,
      xlabel={Environments},
      symbolic x coords ={%
          open,
          suburban,
          intermediate tree-shadow,
          heavy tree-shadow,
          urban,
      },
      xticklabel=\empty,
      ]

      \addplot[
        bar shift=-5pt,fill=white,postaction={
        pattern=north east lines, pattern color=black
    }] coordinates { (open, 0) };
      \addplot[bar shift=5pt,fill=white] coordinates { (open, 0) };

      \addplot[
        bar shift=-5pt,fill=plt1!70,postaction={
        pattern=north east lines, pattern color=black
    }] coordinates { (open, 66.28719931983984) };
      \addplot[bar shift=5pt,fill=plt1!70] coordinates { (open, 67.76665563361574) };

      \addplot[
      bar shift=-5pt,fill=plt2!70,postaction={
        pattern=north east lines, pattern color=black
    }] coordinates { (suburban, 59.97375970276803) };
      \addplot[bar shift=5pt,fill=plt2!70] coordinates { (suburban, 64.41231830717751) };
      
      \addplot[
        bar shift=-5pt,fill=plt3!70,postaction={
        pattern=north east lines, pattern color=black
    }] coordinates { (intermediate tree-shadow, 58.080011031063236) };
      \addplot[bar shift=5pt,fill=plt3!70] coordinates { (intermediate tree-shadow, 63.39092156973394) };
      
      \addplot[
        bar shift=-5pt,fill=plt6!70,postaction={
        pattern=north east lines, pattern color=black
    }] coordinates { (heavy tree-shadow, 57.512479690636944) };
      \addplot[bar shift=5pt,fill=plt6!70] coordinates { (heavy tree-shadow, 60.45771389019385) };
      
      \addplot[
        bar shift=-5pt,fill=plt5!70,postaction={
        pattern=north east lines, pattern color=black
    }] coordinates { (urban, 57.03474187203291) };
      \addplot[bar shift=5pt,fill=plt5!70] coordinates { (urban, 62.0518730036318) };
      
    \end{axis} 
  \end{tikzpicture}
    \caption{Deep shadow, compression ratio 0.33}
    \label{subfig:base4080_deep_shadow_33}
  \end{subfigure}
  
  \vspace{1em}
  
  \centering
  {\hypersetup{hidelinks}\ref{jscc-legendangles}} 
  \tikz{%
  \node[draw=black, inner sep=4.3pt, font=\footnotesize] {
    \tikz{\node[fill=plt1, inner sep=.6ex] {};} open \hspace{1em}
    \tikz{\node[fill=plt2, inner sep=.6ex] {};} suburban \hspace{1em}
    \tikz{\node[fill=plt3, inner sep=.6ex] {};} intermediate tree shadow \hspace{1em}
    \tikz{\node[fill=plt4, inner sep=.6ex] {};} heavy tree shadow \hspace{1em}
    \tikz{\node[fill=plt5, inner sep=.6ex] {};} urban
  };
}

  \caption{\ac{psnr} achieved by \acs{djscc-sat} for different environments, compression rates, and 40\textdegree{} vs. 80\textdegree{} elevation angle.}
  \label{fig:elevation_angles}
\end{figure*}

\Cref{fig:elevation_angles} shows the impact of the satellite's elevation angle by comparing performance at 40\textdegree{} and 80\textdegree{}. While in \ac{los} conditions the difference is negligible, stronger performance variations can be seen for the more challenging shadow and deep shadow states. The results also suggest that moderate compression ratios should be chosen in case of low elevation angles.

\subsection{Performance of Adaptable \acs{djscc} (\acs{adjscc-sat})}
\label{subsec:eval_adjsccsat}

\begin{figure*}
  \begin{subfigure}{.48\linewidth}
    \centering
    \begin{tikzpicture}
    \begin{axis}[
        width=\linewidth,
        height=.6\linewidth,
        xlabel = {Compression ratio},
        ylabel = {PSNR [dB]},
        legend cell align={left},
        ymin = 50,
        ymax = 70,
		legend columns=3,
		legend style={/tikz/every even column/.append style={column sep=1em}},
        ylabel near ticks,
        ylabel style={font=\footnotesize},
        xtick = data,
        x tick label style={
            rotate=0,
            /pgf/number format/fixed,
            /pgf/number format/precision=2,
        },
        x label style={below=0},
        legend to name={legendurban}
        ]

        \addplot[mark=triangle*,plt1, thick] 
        plot 
        table[x expr=\thisrowno{0}, y expr=\thisrowno{1}, col sep=space] {plots/no_AF_40_urban_los_los};
        
        \addplot[mark=triangle*,plt2, thick] 
        plot 
        table[x expr=\thisrowno{0}, y expr=\thisrowno{1}, col sep=space] {plots/no_AF_40_urban_shadow_shadow};

        \addplot[mark=triangle*,plt5, thick] 
        plot 
        table[x expr=\thisrowno{0}, y expr=\thisrowno{1}, col sep=space] {plots/no_AF_40_urban_deep_shadow_deep_shadow};

        \addplot[mark=square*,plt1, thick, dotted] 
        plot 
        table[x expr=\thisrowno{0}, y expr=\thisrowno{1}, col sep=space] {plots/AF_SNR_40_urban_los_los};
        
        \addplot[mark=square*,plt2, thick, dotted] 
        plot 
        table[x expr=\thisrowno{0}, y expr=\thisrowno{1}, col sep=space] {plots/AF_SNR_40_urban_shadow_shadow};

        \addplot[mark=square*,plt5, thick, dotted] 
        plot 
        table[x expr=\thisrowno{0}, y expr=\thisrowno{1}, col sep=space] {plots/AF_SNR_40_urban_deep_shadow_deep_shadow};

        \legend{
            \ac{djscc-sat} / LOS,
            \ac{djscc-sat} / shadow, 
            \ac{djscc-sat} / deep shadow,
            \ac{adjscc-sat} / LOS,
            \ac{adjscc-sat} / shadow,
            \ac{adjscc-sat} / deep shadow
            }

    \end{axis}
\end{tikzpicture}
    \caption{40\textdegree{} elevation angle}
  \end{subfigure}
  \hfill
  \begin{subfigure}{.48\linewidth}
    \centering
    \begin{tikzpicture}
    \begin{axis}[
        width=\linewidth,
        height=.6\linewidth,
        xlabel = {Compression ratio},
        ylabel = {PSNR [dB]},
        ymin = 50,
        ymax = 70,
        ylabel near ticks,
        ylabel style={font=\footnotesize},
        xtick = data,
        x tick label style={
            rotate=0,
            /pgf/number format/fixed,
            /pgf/number format/precision=2,
        },
        x label style={below=0},
        ]

        \addplot[mark=triangle*,plt1, thick] 
        plot 
        table[x expr=\thisrowno{0}, y expr=\thisrowno{1}, col sep=space] {plots/no_AF_80_urban_los_los};
        
        \addplot[mark=triangle*,plt2, thick] 
        plot 
        table[x expr=\thisrowno{0}, y expr=\thisrowno{1}, col sep=space] {plots/no_AF_80_urban_shadow_shadow};

        \addplot[mark=triangle*,plt5, thick] 
        plot 
        table[x expr=\thisrowno{0}, y expr=\thisrowno{1}, col sep=space] {plots/no_AF_80_urban_deep_shadow_deep_shadow};

        \addplot[mark=square*,plt1, thick, dotted] 
        plot 
        table[x expr=\thisrowno{0}, y expr=\thisrowno{1}, col sep=space] {plots/AF_SNR_80_urban_los_los};
        
        \addplot[mark=square*,plt2, thick, dotted] 
        plot 
        table[x expr=\thisrowno{0}, y expr=\thisrowno{1}, col sep=space] {plots/AF_SNR_80_urban_shadow_shadow};

        \addplot[mark=square*,plt5, thick, dotted] 
        plot 
        table[x expr=\thisrowno{0}, y expr=\thisrowno{1}, col sep=space] {plots/AF_SNR_80_urban_deep_shadow_deep_shadow};

    \end{axis}
\end{tikzpicture}
    \caption{80\textdegree{} elevation angle}
  \end{subfigure}

  \vspace{1em}
  
  \centering
  {\hypersetup{hidelinks}\ref{legendurban}} 
  \caption{\ac{psnr} of \acs{adjscc-sat} vs. \acs{djscc-sat} in an urban environment.}
  \label{fig:af_vs_no_af}
\end{figure*}
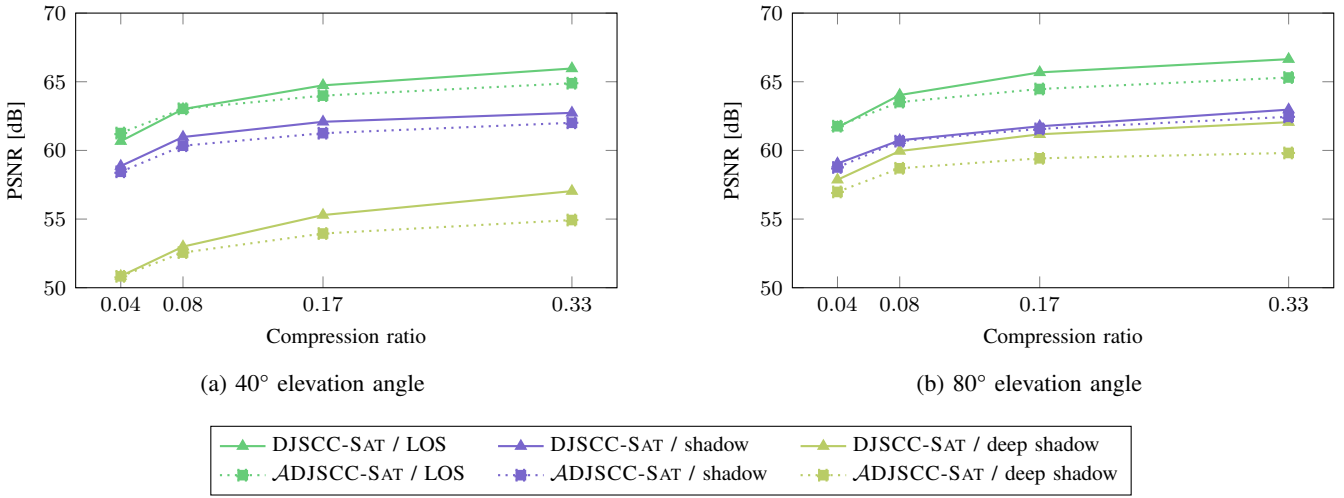

Next, we evaluate how the use of attention modules influences performance by comparing \acs{adjscc-sat} against the \acs{djscc-sat} baseline. For this comparison, we choose an urban environment, as it shows strong performance variability for different states and elevation angles. The results are presented in \Cref{fig:af_vs_no_af}.

\begin{figure*}[t]
  \begin{subfigure}{.5\linewidth}
    \centering
    \begin{tikzpicture}
    \begin{axis}[
        width=\linewidth,
        height=.6\linewidth,
        xlabel = {SNR},
        ylabel = {PSNR [dB]},
        legend cell align={left},
        ymin = 30,
        ymax = 70,
		legend columns=3,
		legend style={/tikz/every even column/.append style={column sep=1em}},
        xtick={4, 8, ..., 20},
        minor xtick={6, 10, ..., 18},
        x tick label style={
            rotate=0,
            /pgf/number format/fixed,
            /pgf/number format/precision=2,
        },
        legend to name={legendurban2}
        ]

        \addplot[mark=triangle*,plt1, thick] 
        plot 
        table[x expr=\thisrowno{0}, y expr=\thisrowno{1}, col sep=space] {plots/no_AF_wrong_snr_0.04_urban_los_40};
        
        \addplot[mark=triangle*,plt2, thick] 
        plot 
        table[x expr=\thisrowno{0}, y expr=\thisrowno{1}, col sep=space] {plots/no_AF_wrong_snr_0.04_urban_shadow_40};

        \addplot[mark=triangle*,plt5, thick] 
        plot 
        table[x expr=\thisrowno{0}, y expr=\thisrowno{1}, col sep=space] {plots/no_AF_wrong_snr_0.04_urban_deep_shadow_40};

        \addplot[mark=square*,plt1, thick, dotted] 
        plot 
        table[x expr=\thisrowno{0}, y expr=\thisrowno{1}, col sep=space] {plots/AF_SNR_wrong_snr_0.04_urban_los_40};
        
        \addplot[mark=square*,plt2, thick, dotted] 
        plot 
        table[x expr=\thisrowno{0}, y expr=\thisrowno{1}, col sep=space] {plots/AF_SNR_wrong_snr_0.04_urban_shadow_40};

        \addplot[mark=square*,plt5, thick, dotted] 
        plot 
        table[x expr=\thisrowno{0}, y expr=\thisrowno{1}, col sep=space] {plots/AF_SNR_wrong_snr_0.04_urban_deep_shadow_40};

        \legend{
            \ac{djscc-sat}{} / LOS,
            \ac{djscc-sat}{} / shadow, 
            \ac{djscc-sat}{} / deep shadow,
            \ac{adjscc-sat}{} / LOS,
            \ac{adjscc-sat}{} / shadow,
            \ac{adjscc-sat}{} / deep shadow
            }

    \end{axis}
\end{tikzpicture}
    \caption{40\textdegree{}, compression ratio $0.04$}
    \label{fig:af_vs_no_af_snr_40_0.04}
  \end{subfigure}%
  \hfill%
  \begin{subfigure}{.5\linewidth}
    \centering
    \begin{tikzpicture}
    \begin{axis}[
        width=\linewidth,
        height=.6\linewidth,
        xlabel = {SNR},
        ylabel = {PSNR [dB]},
        legend cell align={left},
        ymin = 30,
        ymax = 70,
		legend columns=3,
		legend style={/tikz/every even column/.append style={column sep=1em}},
        xtick={4, 8, ..., 20},
        minor xtick={6, 10, ..., 18},
        x tick label style={
            rotate=0,
            /pgf/number format/fixed,
            /pgf/number format/precision=2,
        },
        ]

        \addplot[mark=triangle*,plt1, thick] 
        plot 
        table[x expr=\thisrowno{0}, y expr=\thisrowno{1}, col sep=space] {plots/no_AF_wrong_snr_0.04_urban_los_80};
        
        \addplot[mark=triangle*,plt2, thick] 
        plot 
        table[x expr=\thisrowno{0}, y expr=\thisrowno{1}, col sep=space] {plots/no_AF_wrong_snr_0.04_urban_shadow_80};

        \addplot[mark=triangle*,plt5, thick] 
        plot 
        table[x expr=\thisrowno{0}, y expr=\thisrowno{1}, col sep=space] {plots/no_AF_wrong_snr_0.04_urban_deep_shadow_80};

        \addplot[mark=square*,plt1, thick, dotted] 
        plot 
        table[x expr=\thisrowno{0}, y expr=\thisrowno{1}, col sep=space] {plots/AF_SNR_wrong_snr_0.04_urban_los_80};
        
        \addplot[mark=square*,plt2, thick, dotted] 
        plot 
        table[x expr=\thisrowno{0}, y expr=\thisrowno{1}, col sep=space] {plots/AF_SNR_wrong_snr_0.04_urban_shadow_80};

        \addplot[mark=square*,plt5, thick, dotted] 
        plot 
        table[x expr=\thisrowno{0}, y expr=\thisrowno{1}, col sep=space] {plots/AF_SNR_wrong_snr_0.04_urban_deep_shadow_80};

    \end{axis}
\end{tikzpicture}
    \caption{80\textdegree{}, compression ratio $0.04$}
    \label{fig:af_vs_no_af_snr_80_0.04}
  \end{subfigure}

  \vspace{1em}
  
  \begin{subfigure}{.5\linewidth}
    \centering
    \begin{tikzpicture}
    \begin{axis}[
        width=\linewidth,
        height=.6\linewidth,
        xlabel = {SNR},
        ylabel = {PSNR [dB]},
        legend cell align={left},
        ymin = 30,
        ymax = 70,
		legend columns=3,
        transpose legend,
		legend style={/tikz/every even column/.append style={column sep=1em}},
        xtick={4, 8, ..., 20},
        minor xtick={6, 10, ..., 18},
        x tick label style={
            rotate=0,
            /pgf/number format/fixed,
            /pgf/number format/precision=2,
        },
        x label style={below=0},
        legend to name={legendurban3}
        ]

        \addplot[mark=triangle*,plt1, thick] 
        plot 
        table[x expr=\thisrowno{0}, y expr=\thisrowno{1}, col sep=space] {plots/no_AF_wrong_snr_0.33_urban_los_40};
        
        \addplot[mark=triangle*,plt2, thick] 
        plot 
        table[x expr=\thisrowno{0}, y expr=\thisrowno{1}, col sep=space] {plots/no_AF_wrong_snr_0.33_urban_shadow_40};

        \addplot[mark=triangle*,plt5, thick] 
        plot 
        table[x expr=\thisrowno{0}, y expr=\thisrowno{1}, col sep=space] {plots/no_AF_wrong_snr_0.33_urban_deep_shadow_40};

        \addplot[mark=square*,plt1, thick, dotted] 
        plot 
        table[x expr=\thisrowno{0}, y expr=\thisrowno{1}, col sep=space] {plots/AF_SNR_wrong_snr_0.33_urban_los_40};
        
        \addplot[mark=square*,plt2, thick, dotted] 
        plot 
        table[x expr=\thisrowno{0}, y expr=\thisrowno{1}, col sep=space] {plots/AF_SNR_wrong_snr_0.33_urban_shadow_40};

        \addplot[mark=square*,plt5, thick, dotted] 
        plot 
        table[x expr=\thisrowno{0}, y expr=\thisrowno{1}, col sep=space] {plots/AF_SNR_wrong_snr_0.33_urban_deep_shadow_40};

        \legend{
            \ac{djscc-sat}{} / LOS,
            \ac{djscc-sat}{} / shadow, 
            \ac{djscc-sat}{} / deep shadow,
            \ac{adjscc-sat}{} / LOS,
            \ac{adjscc-sat}{} / shadow,
            \ac{adjscc-sat}{} / deep shadow
            }

    \end{axis}
\end{tikzpicture}
    \caption{40\textdegree{}, compression ratio 0.33}
    \label{fig:af_vs_no_af_snr_40_0.33}
  \end{subfigure}%
  \hfill%
  \begin{subfigure}{.5\linewidth}
    \centering
    \begin{tikzpicture}
    \begin{axis}[
        width=\linewidth,
        height=.6\linewidth,
        xlabel = {SNR},
        ylabel = {PSNR [dB]},
        legend cell align={left},
        ymin = 30,
        ymax = 70,
		legend columns=3,
		legend style={/tikz/every even column/.append style={column sep=1em}},
        xtick={4, 8, ..., 20},
        minor xtick={6, 10, ..., 18},
        x tick label style={
            rotate=0,
            /pgf/number format/fixed,
            /pgf/number format/precision=2,
        },
        x label style={below=0},
        ]

        \addplot[mark=triangle*,plt1, thick] 
        plot 
        table[x expr=\thisrowno{0}, y expr=\thisrowno{1}, col sep=space] {plots/no_AF_wrong_snr_0.33_urban_los_80};
        
        \addplot[mark=triangle*,plt2, thick] 
        plot 
        table[x expr=\thisrowno{0}, y expr=\thisrowno{1}, col sep=space] {plots/no_AF_wrong_snr_0.33_urban_shadow_80};

        \addplot[mark=triangle*,plt5, thick] 
        plot 
        table[x expr=\thisrowno{0}, y expr=\thisrowno{1}, col sep=space] {plots/no_AF_wrong_snr_0.33_urban_deep_shadow_80};

        \addplot[mark=square*,plt1, thick, dotted] 
        plot 
        table[x expr=\thisrowno{0}, y expr=\thisrowno{1}, col sep=space] {plots/AF_SNR_wrong_snr_0.33_urban_los_80};
        
        \addplot[mark=square*,plt2, thick, dotted] 
        plot 
        table[x expr=\thisrowno{0}, y expr=\thisrowno{1}, col sep=space] {plots/AF_SNR_wrong_snr_0.33_urban_shadow_80};

        \addplot[mark=square*,plt5, thick, dotted] 
        plot 
        table[x expr=\thisrowno{0}, y expr=\thisrowno{1}, col sep=space] {plots/AF_SNR_wrong_snr_0.33_urban_deep_shadow_80};

    \end{axis}
\end{tikzpicture}
    \caption{80\textdegree{}, compression ratio 0.33}
    \label{fig:af_vs_no_af_snr_80_0.33}
  \end{subfigure}

  \vspace{2em}
  
  \centering
  {\hypersetup{hidelinks}\ref{legendurban2}} 
  \caption{\ac{psnr} of \acs{adjscc-sat} vs. \acs{djscc-sat} with false \ac{snr} estimation.}
  \label{fig:af_vs_no_af_snr}
\end{figure*}
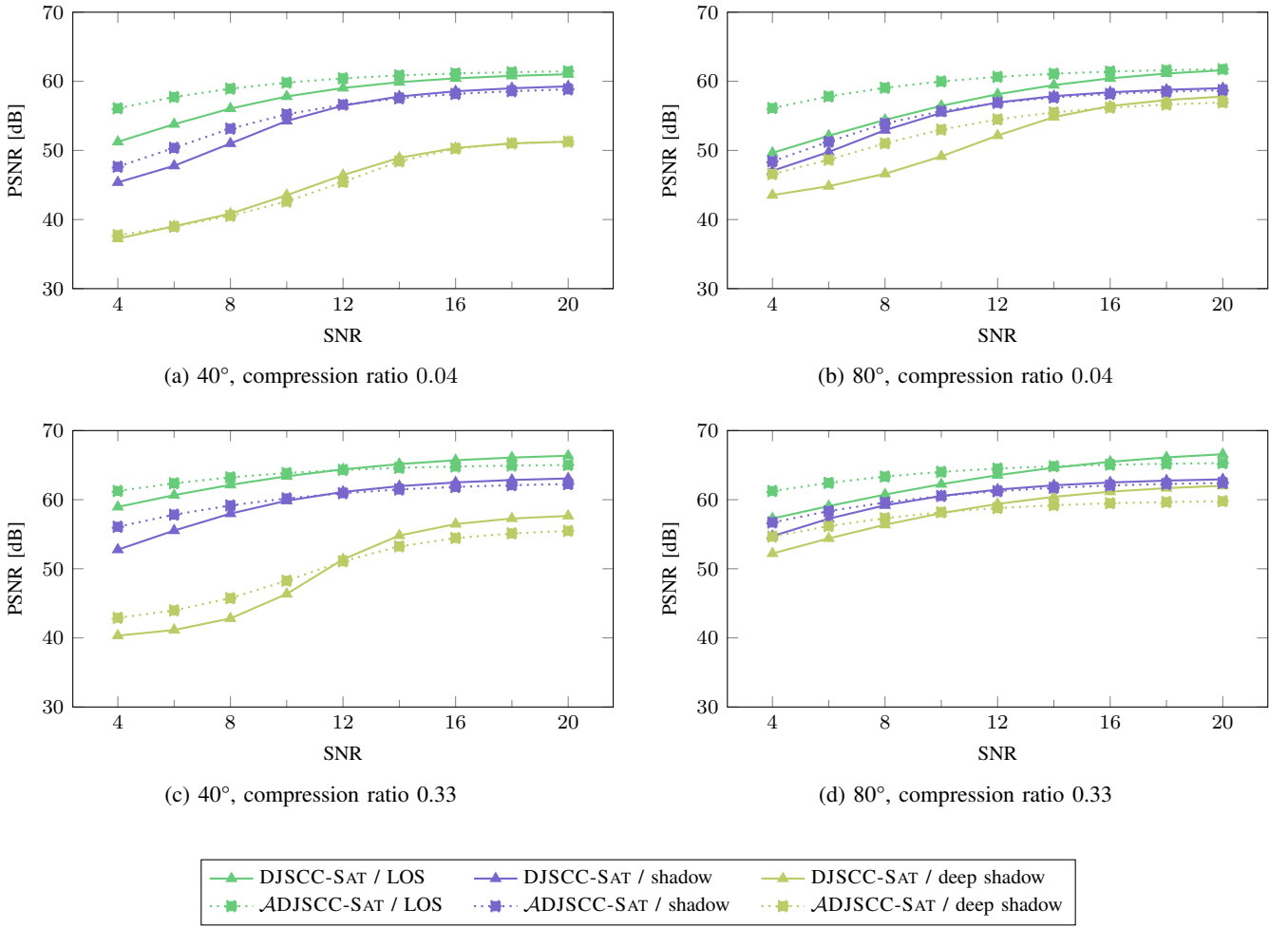

\begin{figure*}[t]
  \begin{subfigure}{.48\linewidth}
    \centering
    \begin{tikzpicture}
    \begin{axis}[
        width=\linewidth,
        height=.75\linewidth,
        xlabel = {Compression ratio},
        ylabel = {PSNR [dB]},
        legend cell align={left},
        ymin = 25,
        ymax = 75,
        ytick distance = 10,
		legend columns=4,
		legend style={/tikz/every even column/.append style={column sep=1em}},
        ylabel near ticks,
        ylabel style={font=\footnotesize},
        xtick = data,
        x tick label style={
            rotate=0,
            /pgf/number format/fixed,
            /pgf/number format/precision=2,
        },
        x label style={below=0},
        legend to name={legenderror}
        ]

        \addplot[mark=triangle*,plt1, thick] 
        plot 
        table[x expr=\thisrowno{0}, y expr=\thisrowno{1}, col sep=space] {plots/no_AF_40_urban_los_los};
        
        \addplot[mark=triangle*,mark options=solid,plt1, thick, dashed] 
        plot 
        table[x expr=\thisrowno{0}, y expr=\thisrowno{1}, col sep=space] {plots/no_AF_40_40_urban_deep_shadow_los};

        \addplot[mark=triangle*,plt2, thick] 
        plot 
        table[x expr=\thisrowno{0}, y expr=\thisrowno{1}, col sep=space] {plots/no_AF_80_urban_los_los};

        \addplot[mark=triangle*,mark options=solid,plt2, thick, dashed] 
        plot 
        table[x expr=\thisrowno{0}, y expr=\thisrowno{1}, col sep=space] {plots/no_AF_80_80_urban_deep_shadow_los};

        \addplot[mark=square*,plt3, thick] 
        plot 
        table[x expr=\thisrowno{0}, y expr=\thisrowno{1}, col sep=space] {plots/AF_SNR_40_urban_los_los};

        \addplot[mark=square*,mark options=solid,plt3, thick, dashed] 
        plot 
        table[x expr=\thisrowno{0}, y expr=\thisrowno{1}, col sep=space] {plots/AF_SNR_40_40_urban_deep_shadow_los};

        \addplot[mark=square*,plt4, thick] 
        plot 
        table[x expr=\thisrowno{0}, y expr=\thisrowno{1}, col sep=space] {plots/AF_SNR_80_urban_los_los};

        \addplot[mark=square*,mark options=solid,plt4, thick, dashed] 
        plot 
        table[x expr=\thisrowno{0}, y expr=\thisrowno{1}, col sep=space] {plots/AF_SNR_80_80_urban_deep_shadow_los};

        \legend{
            \ac{djscc-sat} / correct / 40\textdegree,
            \ac{djscc-sat} / error / 40\textdegree, 
            \ac{djscc-sat} / correct / 80\textdegree,
            \ac{djscc-sat} / error / 80\textdegree,
            \ac{adjscc-sat} / correct / 40\textdegree,
            \ac{adjscc-sat} / error / 40\textdegree, 
            \ac{adjscc-sat} / correct / 80\textdegree,
            \ac{adjscc-sat} / error / 80\textdegree,
            }

    \end{axis}
\end{tikzpicture}
    \caption{LOS instead of deep shadow (better than expected)}
    \label{fig:different_state_same_snr_better}
  \end{subfigure}%
  \hfill%
  \begin{subfigure}{.48\linewidth}
    \centering
    \begin{tikzpicture}
    \begin{axis}[
        width=\linewidth,
        height=.75\linewidth,
        xlabel = {Compression ratio},
        ylabel = {PSNR [dB]},
        ymin = 25,
        ymax = 75,
        ytick distance = 10,
        ylabel near ticks,
        ylabel style={font=\footnotesize},
        xtick = data,
        x tick label style={
            rotate=0,
            /pgf/number format/fixed,
            /pgf/number format/precision=2,
        },
        x label style={below=0},
        ]

        \addplot[mark=triangle*,plt1, thick] 
        plot 
        table[x expr=\thisrowno{0}, y expr=\thisrowno{1}, col sep=space] {plots/no_AF_40_urban_deep_shadow_deep_shadow};
        
        \addplot[mark=triangle*,mark options=solid,plt1, thick, dashed] 
        plot 
        table[x expr=\thisrowno{0}, y expr=\thisrowno{1}, col sep=space] {plots/no_AF_40_40_urban_los_deep_shadow};

        \addplot[mark=triangle*,plt2, thick] 
        plot 
        table[x expr=\thisrowno{0}, y expr=\thisrowno{1}, col sep=space] {plots/no_AF_80_urban_deep_shadow_deep_shadow};

        \addplot[mark=triangle*,mark options=solid,plt2, thick, dashed] 
        plot 
        table[x expr=\thisrowno{0}, y expr=\thisrowno{1}, col sep=space] {plots/no_AF_80_80_urban_los_deep_shadow};

        \addplot[mark=square*,plt3, thick] 
        plot 
        table[x expr=\thisrowno{0}, y expr=\thisrowno{1}, col sep=space] {plots/AF_SNR_40_urban_deep_shadow_deep_shadow};

        \addplot[mark=square*,plt3,mark options=solid, thick, dashed] 
        plot 
        table[x expr=\thisrowno{0}, y expr=\thisrowno{1}, col sep=space] {plots/AF_SNR_40_40_urban_los_deep_shadow};

        \addplot[mark=square*,plt4, thick] 
        plot 
        table[x expr=\thisrowno{0}, y expr=\thisrowno{1}, col sep=space] {plots/AF_SNR_80_urban_deep_shadow_deep_shadow};

        \addplot[mark=square*,plt4,mark options=solid, thick, dashed] 
        plot 
        table[x expr=\thisrowno{0}, y expr=\thisrowno{1}, col sep=space] {plots/AF_SNR_80_80_urban_los_deep_shadow};

    \end{axis}
\end{tikzpicture}
    \caption{Deep shadow instead of LOS (worse than expected)}
    \label{fig:different_state_same_snr_worse}
  \end{subfigure}

  \vspace{2em}
  
  \centering
  {\hypersetup{hidelinks}\ref{legenderror}} 
  \caption{Channel conditions differing from the expected state in an urban environment.}
  \label{fig:different_state_same_snr}
\end{figure*}

The performance of both models depends on the compression ratio and the channel state, while the elevation angle plays a less important role. When stronger compression is applied, the \ac{psnr} values of \acs{adjscc-sat} are similar to that of \acs{djscc-sat}, which uses separate neural networks. This demonstrates that a single adaptable network can achieve comparable performance to multiple specialized networks, while significantly reducing storage overhead (the attention module parameters comprise only $0.25\%$ of all model parameters). For less aggressive compression, \acs{adjscc-sat} shows slightly inferior results. The biggest performance gap is observed in deep shadow conditions and the smallest in shadow conditions. This gap becomes even smaller for shadow conditions when the elevation angle is set to 80°. However, there is no significant difference between 40° and 80° elevation for the other channel states.

\subsection{Robustness to Channel Estimation Errors}
\label{subsec:eval_mismatch}

To conclude our evaluation, we assess the performance of both architectures under channel estimation errors. Our robustness analysis investigates two distinct mismatch scenarios: first, where only the \ac{snr} is incorrectly estimated, and second, where the channel state (defined by $\alpha$, $\psi$, and \MP) is misidentified while the \ac{snr} value remains fixed to its training-time setting.

\Cref{fig:af_vs_no_af_snr} presents the results for the \ac{snr} mismatch scenario. The models, which were trained at \ac{snr} values of 17.25 dB (for 40°) and 20.38 dB (for 80°), were evaluated across a \ac{snr} range of 4 to 20 dB. At a high compression ratio of 0.04 (Figures \ref{fig:af_vs_no_af_snr_40_0.04} and \ref{fig:af_vs_no_af_snr_80_0.04}), performance predictably degrades at lower \acp{snr}. Here, \acs{adjscc-sat} shows superior robustness, outperforming the baseline at low \acp{snr} while matching it at higher values. This indicates that the adaptable model's training across multiple configurations enhances its resilience to large parameter deviations.

At a lower compression ratio of 0.33 (Figures \ref{fig:af_vs_no_af_snr_40_0.33} and \ref{fig:af_vs_no_af_snr_80_0.33}), overall performance improves. The trend of \acs{adjscc-sat} outperforming the baseline at lower \acp{snr} continues. However, for higher \acp{snr} that are closer to the training values, the specialized \acs{djscc-sat} model performs marginally better, consistent with the earlier findings in \Cref{subsec:eval_adjsccsat}.

\Cref{fig:different_state_same_snr} illustrates the second scenario, where the channel state is misidentified. We present results comparing performance with correct channel estimation (solid lines) and incorrect channel estimation (dashed lines). We first analyze a ``better-than-expected'' case, where the system is configured for deep shadow but operates in \ac{los} (\Cref{fig:different_state_same_snr_better}). Interestingly, neither architecture is able to leverage the improved channel conditions, and both show performance degradation. In this case, performance varies with elevation angle: while both models perform similarly at 80°, \acs{adjscc-sat} has a significant advantage at 40°.

Conversely, in the ``worse-than-expected'' case -- where the system anticipates \ac{los} but encounters deep shadow (\Cref{fig:different_state_same_snr_worse}) -- performance degrades severely for both models. However, \acs{adjscc-sat} consistently outperforms the baseline across both elevation angles. This result highlights an important advantage of the adaptable architecture: beyond storage efficiency, its attention modules provide enhanced robustness when channel state estimations are incorrect.

\section{Conclusion}
\label{sec:conclusion}

Small satellite missions, particularly those focused on Earth observation, can produce vast amounts of high-resolution data, but their limited transmission capabilities create a significant communication bottleneck. While separate source and channel coding are traditionally used, their theoretical optimality does not always hold in practice, especially under the harsh and dynamic channel conditions typical for space scenarios.

In this paper, we presented a comprehensive \ac{djscc} framework tailored for small satellite applications. We first designed a basic \ac{djscc} system (\acs{djscc-sat}) and showed its effectiveness for transmitting multi-spectral satellite imagery. To enhance this for practical deployment, we integrated a realistic statistical channel model that accounts for varying channel conditions. To address the complexity and overhead of using multiple models for these diverse conditions, we further proposed an advanced, adaptable architecture (\ac{adjscc-sat}) that augments the \ac{djscc} network with attention modules.

The efficacy of our proposed framework is demonstrated through extensive evaluations using realistic Sentinel-2 data and channel model. The results show that a single, compact network equipped with attention modules can be parameterized for diverse channel conditions, yielding performance on par with multiple, separately trained networks while offering significant storage savings. A key finding is that this adaptability also leads to increased robustness, as the single model outperforms specialized ones in scenarios with channel estimation errors.

This research underscores the value of adaptable \ac{djscc} in creating robust and flexible communication protocols for resource-constrained satellites. As for future work, promising directions include extending the framework to complex multi-satellite scenarios for constellation-wide optimization and adapting it to different types of neural networks for handling non-vision-based data.

\section*{Acknowledgements}

This work has been partially funded by the LOEWE initiative (Hessen, Germany) within the emergenCITY center [LOEWE/1/12/519/03/05.001(0016)/72].


\end{document}